\documentclass{PoS}
\usepackage{epsfig,epsf}
\title{Is the ejection of the corona a general phenomenon in microquasars?}

\ShortTitle{Coronal ejections in microquasars}

\author{\speaker{J\'er\^ome Rodriguez}\\
        CEA Saclay, IRFU/Service D'Astrophysique, UMR AIM, France\\
        E-mail: \email{jrodriguez@cea.fr}}

\author{Lionel Prat\\
        CEA Saclay, IRFU/Service D'Astrophysique, UMR AIM, France\\
        E-mail: \email{lionel.prat@cea.fr}}

\abstract{We study the evolution of some microquasars during their outbursts as 
observed with the X-ray telescopes RXTE and INTEGRAL. We 
focus on the interplay between the accretion disc, and the medium
responsible for the production of the hard X-rays (the so-called corona).  
By comparing the behaviour of two sources (XTE J1550-564 and 
GRS 1915+105) at X-ray energies and radio wavelengths, we 
propose a scenario in which the  discrete ejections are triggered in coincidence 
with soft X-ray peaks during the outburst. 
We also suggest, in those two sources,  that the ejected material is 
the corona that is seen to disappear in coincidence with the X-ray maxima. 
We then turn to two other sources, XTE J1748$-$248, and XTE J1859+226,  
and study whether the same conclusions can be drawn from the existing multi-wavelength 
(radio+X-ray) data.}

\FullConference{7th INTEGRAL Workshop\\
		 September 8-11 2008\\
		 Copenhagen, Denmark}

\begin{document}

\section{Introduction}
\begin{figure}
\caption{Typical schematic evolution of a microquasar through an outburst. The central 
panel shows the hardness-intensity diagram obtained from the X-ray observations, with, on 
each sides, the corresponding behaviour of the accretion disc and the jet. The outer panels 
show the corresponding behaviour in the Fourier domain. Adapted from 
\cite{fender05,mclintock06,homan06}.}
\label{fig:q-shape}
\epsfig{file=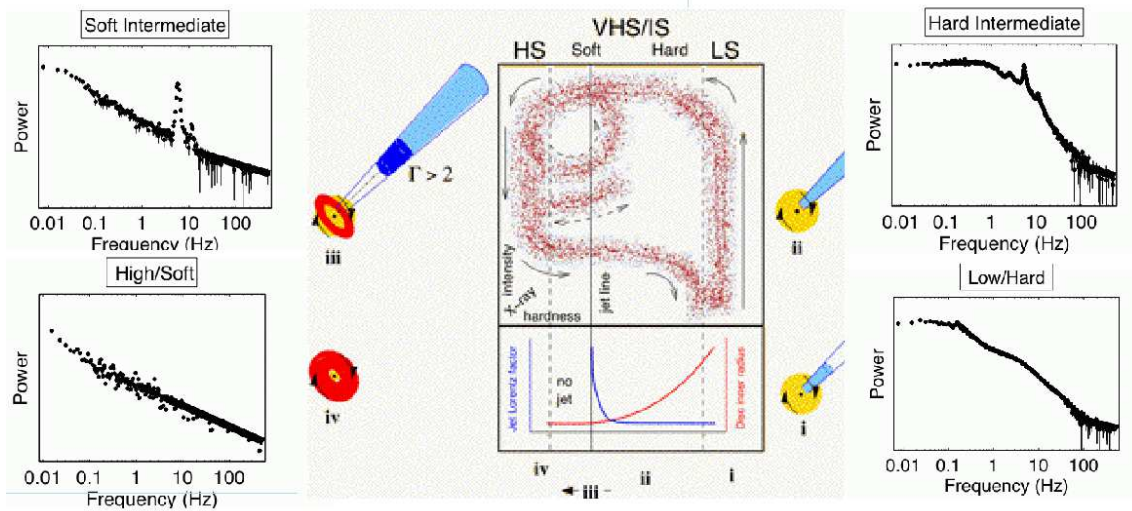, width=\columnwidth}
\end{figure}

Up to now, about 20 black hole (BH) X-ray transients have been observed, along with 20 more 
candidates black-hole binaries \cite{mclintock06}. The huge archival X-ray observations 
available allowed different spectral states to be identified. These depend on both the 
spectral and temporal characteristics of the source \cite{homan06,mclintock06}. Radio 
observations have shown that these objects, through their outbursts, usually show the 
presence of radio jets. This behaviour, also seen in quasars that are thought to host 
super-massive BH, led, by analogy, to call them 'microquasars' \cite{mirabel94}. Since then, 
systematic quasi-simultaneous 
multi-wavelengths observations (i.e. combining, at least, radio and X-ray observations), have
permitted researchers to identify a common pattern for the evolution of microquasars through 
their outbursts, referred to as the ``Q-shape'' given the shape it draws on a hardness-intensity 
diagram (HID, Fig.~\ref{fig:q-shape}, \cite{fender05}).
This approach has the advantage of being model independent. It is, however, interpreted in terms 
of a generally admitted model, involving an accretion disc, responsible for the emission of soft
X-rays, a so-called ``corona'' responsible for the hard X-ray tail, and a jet responsible for 
the radio and part of the infrared emissions. Note that there is no consensus as to whether 
the corona and the jet are different media, and some models predict that the jet could be 
responsible of (part of) the hard X-ray emission \cite{markoff03}. \\
\indent An outburst starts in the ``low'' hard state (LHS). The spectrum is hard, dominated by the 
coronal emission (inverse Comptonization on a population of thermal electrons). The X-rays show
a high level of variability and some quasi-periodic oscillations (QPOs), while a compact jet
is seen in the radio bands. In this state the radio and soft X-ray fluxes are known to be correlated 
\cite{corbel00}. During the outburst, the source may reach the soft state (SS),  
in which the hard tail is almost completely absent, and the X-rays are entirely dominated by  
the disc emission. There is no jet in this state. The transition from LHS to SS occurs 
through different intermediate states, and through the so-called jet line. 
In this top horizontal branch of the HID (Fig.~\ref{fig:q-shape}), the compact jet is quenched,
while the crossing of the jet line is accompanied by discrete ejections of material \cite{fender05}.
At the end of the outburst, the source returns to quiescence after it has transited again the the LHS.
A key to fully understand the physics of these states may reside in deeper and more systematic 
studies of the state transitions, especially with physical diagnostics. \\
\indent In this paper we initiate a systematic X-ray study of microquasars in dates preceding
radio ejections. This follows earlier works by \cite{rodriguez03} and \cite{rodriguez08a,rodriguez08b}
on, respectively, XTE J1550$-$564 and GRS 1915+105, who showed in particular similar evolution at
times preceding ejections. \\
\indent We start this paper by summarising the results obtained in XTE J1550$-$564 and GRS 1915+105 in 
Section 2 and 3, and then turn to two other sources, XTE J1748$-$248, and XTE J1859+226, for which quasi-simultaneous 
radio and X-ray observations exist. We discuss our results in the last part of the paper.

\section{XTE~J1550$-$564: the beginning of the story?}
\begin{figure}
\caption{{\it{Left :}} Evolution of the main spectral parameters of XTE J1550$-$564 during 
the 2000 outburst. {\it{Right :}} Evolution of the 2--50 keV disc flux vs. the 2--50 keV power law 
 flux over the period of outburst. The state prescription is that of \cite{mclintock06}. 
Both figures adapted from \cite{rodriguez03}.}
\label{fig:1550}
\epsfig{file=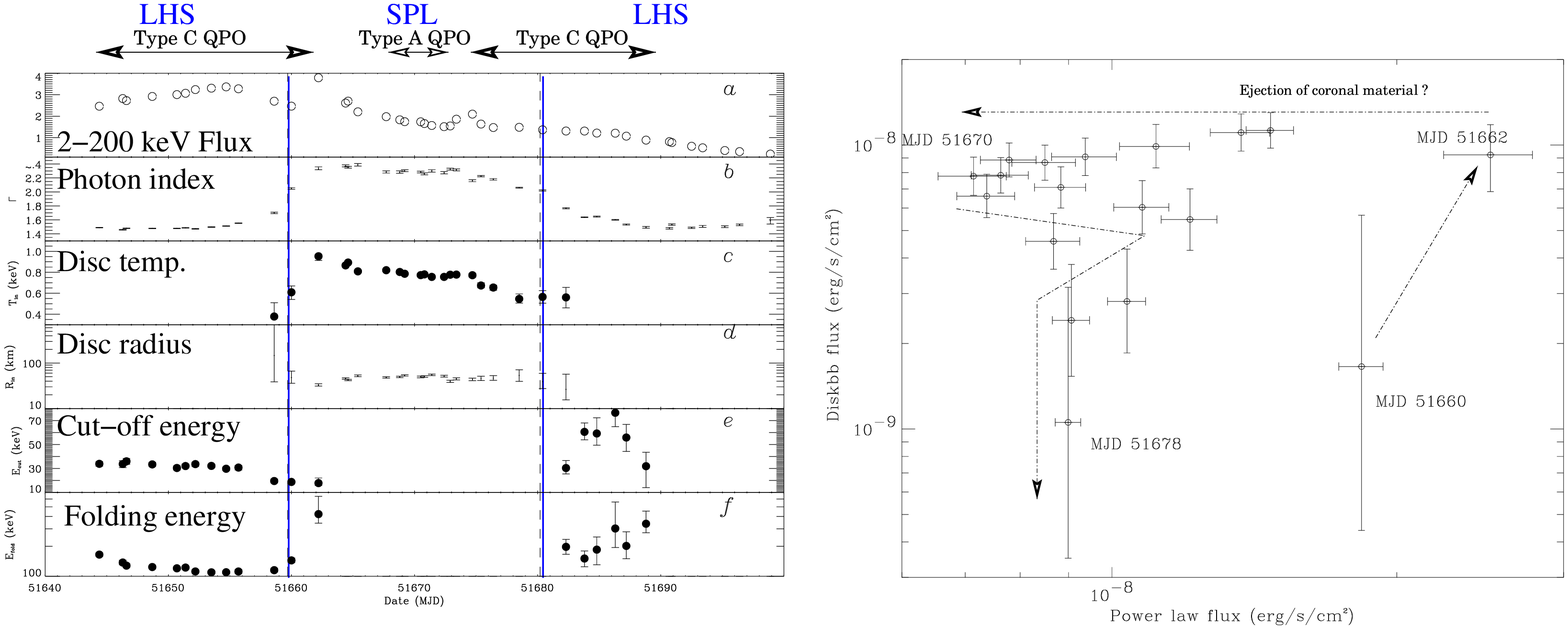, width=\columnwidth}
\end{figure}

The 2000  outburst of XTE J1550$-$564 ($\sim$70 day) was monitored in a multi-wavelength approach. Mainly, this 
included  X-ray energies with RXTE, and radio and infrared wavelengths from ground based facilities. 
\cite{corbel01} reported the detection of radio emission soon after the 2--12 keV maximum of the outburst. 
This radio emission was interpreted as due to a discrete ejection of material,  the precise date of which 
could not be determined. \cite{rodriguez03,rodriguez04} presented a precise spectral and timing 
analysis of all RXTE observations of this outburst. The energy spectra were fitted with the 
common model involving emission from an accretion disc, a power law tail (either with a high energy 
cut-off or not), an iron line and an iron edge around 8 keV. The evolution of the main spectral 
parameters, and the identification of the spectral states (according to the classification of \cite{mclintock06} 
are reported in Fig.~\ref{fig:1550}.
 The cut-off power law is commonly thought to represent a process of thermal Comptonization, that 
occurs in the corona. The absence of a cut-off is usually taken as evidence that the corona  is no more 
thermalised (i.e. the electrons no longer have a Maxwellian distribution of their velocities). 
During the 2000 outburst of XTE J1550$-$564, the cut-off power law disappeared about 2 days 
after the first transition, on MJD 51662, and reappeared after 
the second transition, when the source went back to the LHS, on MJD 51682 (Fig.~\ref{fig:1550}). 
Soon after the X-ray maximum (after MJD 51662), while the disc flux remained approximately 
constant over few days, the power law flux decreased significantly. At the same time the power law 
photon index remained roughly constant, which indicates that the decrease of the power law 
flux is not due to a pivoting of the X-ray spectrum but traces the fact that something drastic changed in  the 
corona. The large decrease of the coronal flux may indicate that part of the corona disappeared, ie, that it 
was either accreted by the BH, or ejected. The detection of a discrete ejecta soon after the 
X-ray peak  makes it tempting to consider that the Compton medium may have been blown away. 

\section{GRS 1915+105: towards a generalisation of the ejection of the corona}
\begin{figure}
\caption{Zoom on a portion of the INTEGRAL and Ryle observations of GRS 1915+105 showing three particular
cycles, each followed by a radio flare. The 3 observations belong to classes $\nu$, $\lambda$, and $\beta$. 
From \cite{rodriguez08b}.}
\label{fig:1915}
\epsfig{file=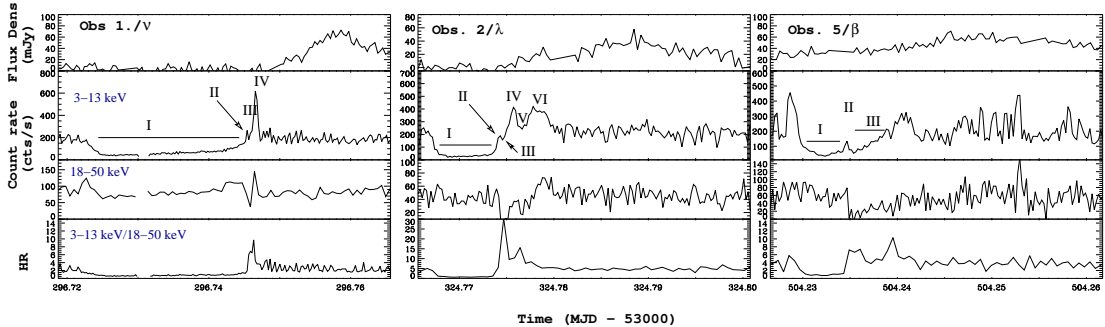,width=\columnwidth}
\end{figure}

\begin{figure}
\caption{Spectral evolution of GRS 1915+105 during the cycle of class $\nu$. The fit-model
consist of disc black body and Comptonised emission both absorbed.}
\label{fig:1915fits}
\epsfig{file=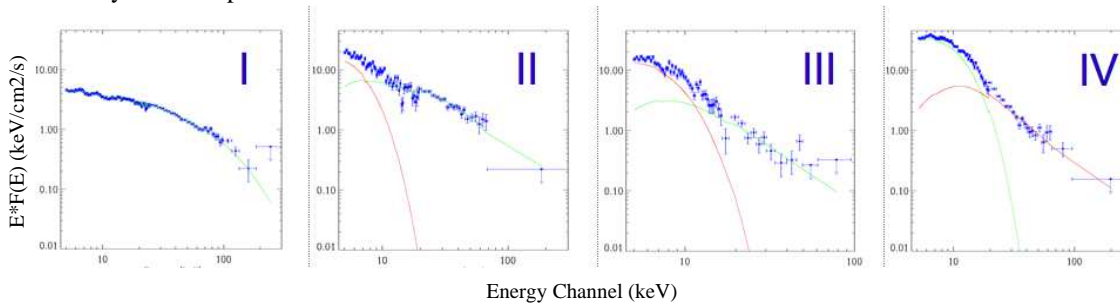,width=\columnwidth}
\end{figure}

GRS~1915+105 has been widely observed at all wavelengths since its discovery in 1992. 
First connection between accretion and ejection phenomena were reported by \cite{mirabel98}, 
when these authors remarked that during in a particular observation, X-ray episodes 
of low and spectrally hard state terminated by a sudden X-ray spike marking the transition 
to a softer state (hereafter ``cycle''), were followed by flares, seen in infrared first and later 
in radio. These were interpreted as due to discrete ejection of plasmoids.\\
\indent We (PIs Hannikainen and 
Rodriguez) are conducting a monitoring campaign of GRS 1915+105 with INTEGRAL since 2003. This campaign is,
as much as possible, conducted in coordination with radio observations. Over the period 2004--2005, 
we were lucky enough to catch the source in several of its variability states (Fig.~\ref{fig:1915}) 
as defined by \cite{belloni00}. So far, amongst all classes showing cycles (the so-called 
$\alpha$, $\beta$, $\nu$, $\lambda$, $\theta$), only the cycles of class $\lambda$ had not 
been seen to be associated to radio flares, and thus to ejections \cite{klein02}.
The observation of an ejection during a $\lambda$ observation was the first ever reported. It has a great 
importance as it allowed us to generalise the following: in GRS 1915+105, discrete plasmoid ejections 
always occurs as a response to an X-ray cycle composed of a spectrally hard X-ray dip (longer than 100s)
ended by a sudden X-ray spike maring the return to a spectrally soft state (Fig.~\ref{fig:1915}, 
\cite{rodriguez08a}). We indeed 
showed that, {\it{from both spectral and temporal points of view}},  theses classes were qualitatively
similar. They all show the presence of low frequency QPOs of variable frequency during the X-ray dip. 
The QPO disappears at the spike. The delay between the X-ray spike and the radio peak is similar 
(about 0.3 hours). The spectral evolution during the cycle is similar (although the exact spectral parameters
may have different absolute values). There exists a possible correlation between the duration of the 
X-ray dip and the amplitude of the radio flare, which may strongly link what happens in the disc-corona
system prior to the ejection \cite{rodriguez08a}.\\
\indent The generalisation of the accretion-ejection links in GRS 1915+105 was shown in two 
different ways, a model-independent one (i.e. based on light curves and hardness ratios \cite{rodriguez08a}, not 
shown here), and through
fits to the spectra (\cite{rodriguez08b}, Fig.~\ref{fig:1915fits}). The cycles were divided into different intervals 
based on the hardness ratio of the source. All spectra were fitted with a model accounting for 
the emission of the disc ({\tt{ezdiskbb}}) and that of the corona ({\tt{comptt}}), both modified by 
absorption. In all cases, the transition from interval II to III manifests by a drastic change in the 
coronal properties. First the corona seems to completely change from thermalised to non-thermalised. 
Then the flux of the corona shrinks while that of the disc remains constant. 
Again, all the points mentioned above, and the spectral evolution of the source through the cycle, 
leads us to suggest that (part of) the corona is ejected \cite{rodriguez08a,rodriguez08b}.\\
\section{What happens in other sources : the cases of XTE J1859+226 and XTE J1748$-$288}
The common behaviour of XTE J1550$-$564 and GRS 1915+105, at least regarding the origin
of the ejected material, led us to search for the same evolutionary pattern in other sources. 
Here we present the results obtained for two more microquasars.
 
\subsection{Evolution of the fluxes: a different picture?}
\begin{figure}
\caption{Multiwavelength evolution of XTE J1748$-$288 (Left) over its 1998 outburst, and XTE J1859+226 (Right) 
over its 1999 outburst. From top to bottom,  
the panels show the radio light curves seen with the GBI, the soft (RXTE) and hard (BATSE)
X-rays, intermediate X-rays as seen with RXTE/PCA, and the hardness ratio between two X-ray bands.}
\label{fig:1748}
\epsfig{file=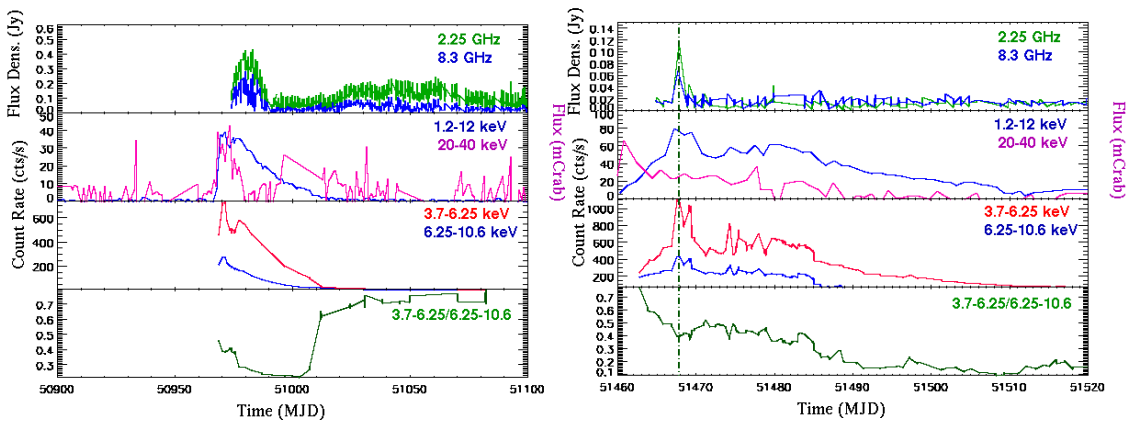,width=\columnwidth}
\end{figure}

XTE J1748$-$288 is another transient source that was discovered by RXTE in 1998. \cite{mike00}
report the spectral analysis of the RXTE data over the outburst. Although they analyse
the data with phenomenological models, they remark that the 15-30 keV flux decreased more rapidly
than the soft X-ray flux after the peak of the outburst.
A multiwavelength light curve of this 1998 outburst, with the Green Bank Interferometer in radio, 
RXTE at X-ray energies, and CGRO/BATSE in the hard X-rays, is shown in the left panel of Fig.~\ref{fig:1748}.\\
\indent XTE J1859+226 was also discovered by RXTE in 1999 \cite{woods99}, when entering an X-ray outburst.
The multiwavelength light curve is shown in Fig~\ref{fig:1748}. A deep radio analysis is presented
by \cite{brocksopp02} who remark that the source showed multiple ejection events during the outburst.
The multiwavelength light curve is shown in the right panel of Fig~\ref{fig:1748}. 
Similarly to XTE~J1550$-$564 and GRS 1915+105 (although on a much longer time scale), the same 
kind of qualitative evolution can be seen in these two sources. The outburst starts at hard X-rays. 
The soft X-ray peak is followed by a radio flare, indicative of an ejection of material.\\

\begin{figure}
\caption{Evolution of the fluxes of the different spectral components (disc in green and Comptonised component
in blue) vs time (top panel).
Evolution of the 2--50 keV disc flux vs. the 2--50 keV Comptonised flux (bottom panel).
The left side of the figure corresponds to the data of XTE J1748$-$288, while the right side
corresponds to XTE J1859+226. The vertical lines represent the dates of the first radio detection and of 
the radio maximum for XTE J1748$-$288, and that of the first detection of an optically thin radio ejection in 
XTE J1859+226.}
\label{fig:fluxes}
\epsfig{file=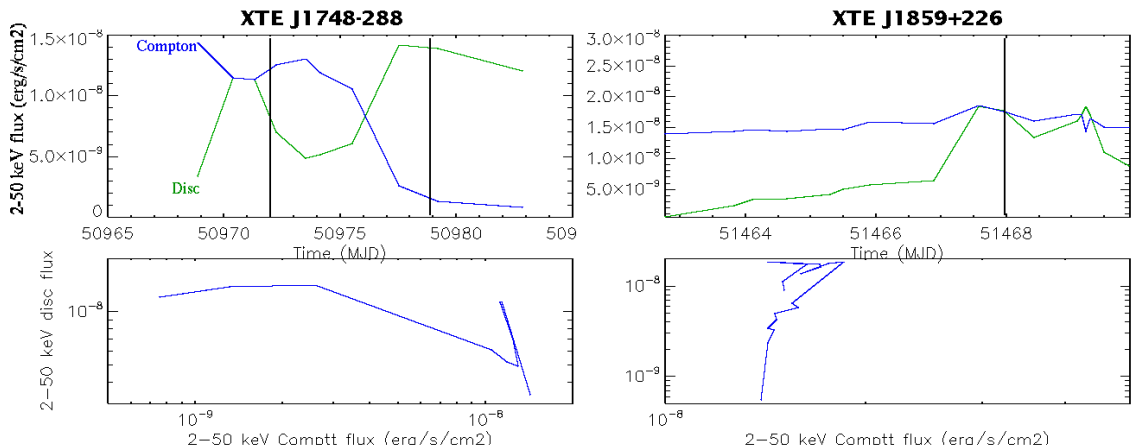,width=\columnwidth}
\end{figure}

We reduced the RXTE (PCA+HEXTE) data of all the observations from the beginning of the 
outburst of the two sources until the peak of the radio emission. The data reduction was performed in a 
very standard way (in particular concerning the time filtering, production of light curves 
and spectra). See, for example, \cite{rodriguez03,rodriguez08b} for the details of the RXTE data 
reduction which are identical. The resulting spectra were fitted in XSPEC with a model consisting of a disc 
component (diskbb) and Comptonization (comptt) to account for the hard X-ray emission. All these
components were convolved by low energy absorption. The choice of this particular model (as opposed to a 
simple power law) is dictated by the fact that the interplay between the disc and the 
corona is more physical at energies $<3$kT$_{disc}$. In particular the model suffers less of the 
mixing of a high level power law in the region where the disc emits at maximum. Therefore, 
the fluxes of the two components are more realistic. In all cases the temperature of the seed 
photons for Comptonization was equalised  to that of the disc. For both sources we therefore obtained 
all spectral parameters: disc temperature, disc inner radius, temperature of the Comptonising electrons, optical
depth of the Compton component, and the unabsorbed fluxes of the two spectral components.  The detailed 
spectral results will be presented in a future paper (Rodriguez \& Prat, in prep.).\\
\indent  Here we focus on the evolution of the fluxes of the two spectral components vs. 
time, and that of the 2--50 keV  disc flux one vs. the 2--50 keV comptonised flux, 
which are represented in  Fig.~\ref{fig:fluxes}. 
In both cases, we also represented significant radio events (beginning and/or maxima) with vertical lines. 
In XTE J1748$-$288 a significant
decrease of the Comptonised flux is seen prior to the radio peak, and this could be indicative, as in the 
previous two sources, of ejection of coronal material. The date of the first radio detection 
(\cite{hjellming98}), however, occured before the decrease was initiated. It even occured after a slight decrease
is seen in the flux of the accretion disc (Fig.~\ref{fig:fluxes}). This therefore seems to contradict
the picture we drew in the cases of XTE J1550$-$564 and GRS 1915+105. The huge decrease in coronal 
flux, however, may still indicate that a significant amount of coronal material feeds the jet.\\
\indent The case of XTE J1859+226 is apparently completely at odds with this interpretation. The 
flux of the Comptonised component remains roughly constant over the period of interest (Fig.~\ref{fig:fluxes}).
This source is even more puzzling as the disc flux also shows an opposite behaviour to what would (naively)
be expected. Indeed, it increases before the radio flare.

\subsection{Do we look at the right parameters?}
\begin{figure}
\caption{Evolution of the $y$ and $\tau$ parameters during the outburst of XTE J1748$-$288 (left) and 
XTE J1859+226 (right). In both cases, the top panels shows the PCA 3.7--6.25 keV light curve. The vertical lines
represent the first radio detection in XTE J1748$-$288, and the first detection of 
optically thin synchrotron emission in XTE J1859+226. }
\label{fig:tau}
\epsfig{file=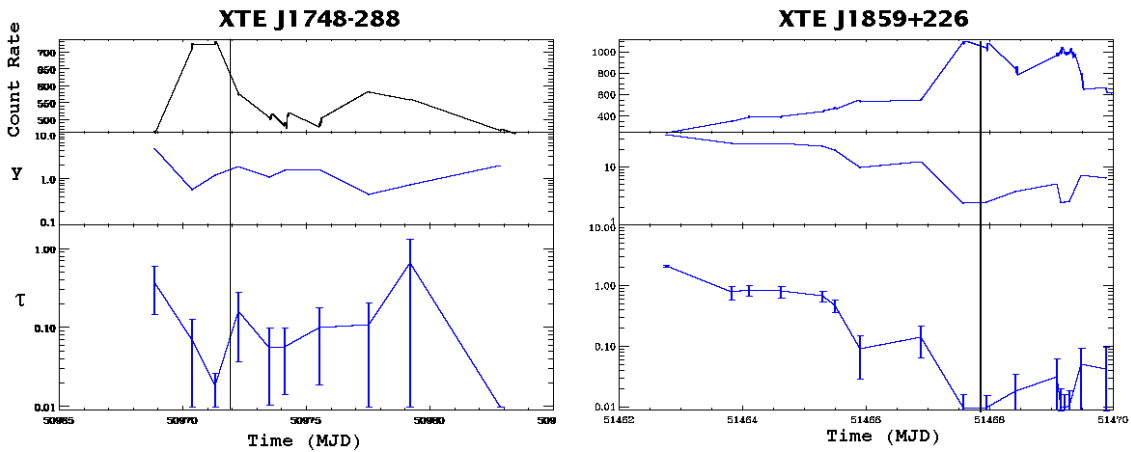,width=\columnwidth}
\end{figure}

We look at the spectral components of the accretion flow and search if any 
of the two undergoes significant evolution. In XTE J1550$-$564 and GRS 1915+105 the drop
in the flux of the Comptonised component led us to conclude that the ejected matter detected 
soon after this decrease was made of coronal material. The picture may, however, not be so simple 
in all sources, as Comptonization and transitions 
from a thermalised corona  to a non thermalised one are not trivial phenomena. The evolution of 
fluxes may not be the appropriate parameter to look at. Indeed, one may think 
of a system in which the amount of seed photons increases such that even with a shrink of 
Comptonising electrons, the hard X-ray flux could appear as not varying much. In fact, the 
right parameter to characterise Comptonization is the so-called $y$ parameter (\cite{rybicki})
that somehow traduces the efficiency of Comptonization. $y$ can be expressed in terms of 
physical quantities characterising the corona, since $y\propto kT_e \times max(\tau,\tau^2)$, 
with $kT_e$ the temperature of the coronal electrons, and $\tau$ the optical depth of the corona. 
$max(\tau,\tau^2)$ is equivalent to the mean number of scattering, and one may further remark 
that $\tau\propto\rho\times R$, with $\rho$ the coronal density and $R$ the size of the corona.
We represented the evolution of the $\tau$ parameters during the outbursts of XTE J1748$-$288 and 
 XTE J1859+226 in Fig.~\ref{fig:tau}. 
While the evolution of the fluxes does not indicate anything special concerning the corona, 
the evolution of the $y$ and $\tau$ parameters shows that something drastic occurred. These quantities
indeed decrease by a factor $\geq10$ in both sources. In each 
source it is remarkable that these drops occurred slightly before the radio ejections appeared.
Recalling that $y$ is somehow related to the mean number of scattering, and that $\tau$ has the dimension of
a surface density, these drops are therefore compatible with: a decrease of the number of coronal electrons (at 
constant coronal size), or a decrease of $R$ with a constant electron density, or even both effects combined.
While the first effect could be due to sudden accretion of the corona, the latter two are 
equivalent to an ejection of the corona, since they would be equivalent to a net loss of mass. 
In any cases, the shrink of $\tau$ indicates that the corona
sees its density decreasing, and that therefore it somehow looses matter. Note that, although we cannot 
rule out that 
a sudden accretion event has taken place, the detection of ejections soon after the 
disappearance of coronal medium again argues in favour of the ejection of part of the corona.

\section{Conclusions}
We have shown that in GRS 1915+105 and XTE J1550$-$564 the origin of 
the ejection could be the coronal material. The argument is based on the evolution of the fluxes
of the different X-ray emitting media. In both cases, the flux of the corona is seen to decrease 
by a large factor before an ejection is detected in the radio domain. 
In order to test this simple model, we studied 
two other sources for which multiwavelength data are available. These sources, XTE J1748$-$288 and 
XTE J1859+226, showed standard outbursts, initiated by a hard X-ray flare, transition to 
a softer outburst, the peak of which preceded a radio flare indicative of a discrete ejection
of matter. Contrary to GRS 1915+105 and XTE J1550$-$564, the fluxes do not evolve in the same simple 
way in those two sources. In XTE J1859+226 we do not even see a drop in the flux of any of the two 
spectral components. \\
\indent We remarked that the right parameters to properly characterise Comptonization is  
the $y$ parameter, and more precisely the optical depth of the corona $\tau$.
When studying those two quantities, we saw that, for both sources, they dropped by a significant 
factor before the radio ejection was detected. Given that $\tau$ somehow relates to the corona 
density and radius, we interpret these observations as evidence that part of the corona is ejected. 
Of course more systematic multiwavelength monitoring of microquasars are necessary to further 
conclude and generalised the fact that the ejecta are composed of coronal material in all sources.
Our approach showed, that although the details of their evolution were different, in all four microquasar
the origin of the discrete ejection is tightly linked to the evolution of the properties of the corona. 
We interpret the coronal evolution, in all four sources, as a disappearance of part of this medium, that 
feeds the ejecta. 

\section*{Acknowledgements}
We acknowledge E. Kuulkers, R. Rothschild, M. del Santo, M. McCollough for useful interactions during the presentation of 
this work at the meeting.  We particularly warmly thank M. Coriat, S. Corbel, and E. Koerding for constant 
stimulating exchanges and discussions, and a careful reading of this paper. This research has made use of 
data obtained through the High Energy Astrophysics Science Archive Center Online Service, provided by 
the NASA/Goddard Space Flight Center. It is also partly based on observations with INTEGRAL, an ESA 
mission with instruments and science data centre funded by ESA 
member states (especially the PI countries: Denmark, France, Germany, Italy, Switzerland, Spain), 
Czech Republic and Poland, and with the participation of Russia and the USA. The Green Bank Interferometer 
was operated by the National Radio Astronomy Observatory for the U.S. Naval Observatory
and the Naval Research laboratory during the time period of these observations.

\end{document}